\documentclass[aps,prb,floatfix,twocolumn,preprintnumbers,amsmath,amssymb,groupedaddress,showpacs,showkeys]{revtex4-2}

\usepackage{graphicx} 
\usepackage{dcolumn} 
\usepackage{bm} 
\usepackage{color} 
\usepackage{amsmath,amssymb}
\usepackage{mathrsfs}
\usepackage{upgreek}
\usepackage{mathtools}

\usepackage[colorlinks=true,linkcolor=blue,citecolor=blue,urlcolor=blue]{hyperref}

\usepackage{xr}
\makeatletter
\newcommand*{\addFileDependency}[1]{
  \typeout{(#1)}
  \@addtofilelist{#1}
  \IfFileExists{#1}{}{\typeout{No file #1.}}
}
\makeatother

\newcommand*{\myexternaldocument}[1]{%
    \externaldocument{#1}%
    \addFileDependency{#1.tex}%
    \addFileDependency{#1.aux}%
}

\myexternaldocument{SM}

\begin{document}

\title{Phenomenological Theory of the Supercurrent Diode Effect: The Lifshitz Invariant}

\author{Denis Kochan$^{1,2}$}\email[Corresponding author: ]{denis.kochan@savba.sk}
\author{Andreas Costa$^2$}
\author{Iaroslav Zhumagulov$^2$}
\author{Igor \v{Z}uti\'{c}$^{3}$}
\affiliation{%
$^1$Institute of Physics, Slovak Academy of Sciences, 84511 Bratislava, Slovakia \\	
$^2$Institut f\"ur Theoretische Physik, Universit\"at Regensburg, 93053 Regensburg, Germany\\
$^3$University at Buffalo, State University of New York, Buffalo, New York 14260-1500, USA 
}%


\begin{abstract}
   Nonreciprocal phenomena in the normal state are well established and key to many commercial applications. In contrast, superconducting analogs, such as the superconducting diode effect (SDE), are only starting to be experimentally explored and pose significant challenges to their theoretical understanding. In this work we put forth a phenomenological picture of the SDE based on the generalized Ginzburg-Landau free energy, which includes a Lifshitz invariant as the hallmark of noncentrosymmetric helical phase of the finite-momentum Cooper pairs. We reveal that such a Lifshitz invariant drives the SDE in quasi-two-dimensional systems in an applied magnetic field and cannot be removed by a gauge transformation, due to the inherently inhomogeneous magnetic response. For a thin film, the SDE scales with the square of its thickness and nonlinearly with the strength of the in-plane magnetic field.  We derive an explicit formula that relates the SDE at small magnetic fields to the strength of Rashba spin-orbit coupling, g-factor, and Fermi energy. For a noncentrosymmetric Josephson junction, we self-consistently obtain generalized anharmonic current-phase relation which support the SDE. The transparency of our approach, which agrees well with experimentally-measured SDE, offers an important method to study nonreciprocal phenomena, central to superconducting spintronics and topological superconductivity.
\end{abstract}

\date{\today}
\maketitle


\paragraph*{\textbf{Introduction:}} 


Nonreciprocal response is ubiquitous to many phenomena in classical and quantum 
physics~\cite{Coulais2017:Nature,Shadrivov_2011:NJP,Caloz:PhysRevApplied}. 
Faraday and Kerr effects were already well known in the nineteenth 
century~\cite{Faraday1846,Kerr1877},
while the nonreciprocity of the semiconducting diodes has enabled many applications in 
opto-electronics and 
spintronics~\cite{Shockley1952,Zutic2004,Waser2012-ee,Tsymbal2019}. 
However, until the last few years, experimental demonstrations of the nonreciprocal phenomena were largely absent from 
superconductivity~\cite{Wakatsuki2017,Itahashi2020}. 
 
Recent reports of the superconducting diode effect (SDE) in noncentrosymmetric superconductors and Josephson junctions 
(JJs)~\cite{Ando2020,Baumgartner2022,Wu2022,Jeon2022,Pal2022,arxiv.2212.13460,Matos-Shabani}, 
have generated a great interest to examine its relevance to other phenomena and their applications~\cite{arxiv.2301.13564,amundsen2022colloquium}, 
as well as to identify an even earlier SDE 
observations~\cite{amundsen2022colloquium,PhysRevLett.126.036802}. 
SDE has nonreciprocal response with the current direction and is often associated with magnetochiral 
anisotropy~\cite{Ando2020,Baumgartner2022,Wu2022,Jeon2022,Pal2022,arxiv.2212.13460,Matos-Shabani}, 
depending on the vector product of the supercurrent and the applied or proximity-induced magnetic field~\cite{PhysRevLett.99.067004}. 

With a growing number of materials platforms that supports 
SDE~\cite{Ando2020,diezmerida2021magnetic,Baumgartner2022,BaumgartnerSI2022,Wu2022,Jeon2022,Pal2022,Bauriedl2022,Turini2022,lin2022zerofield,arxiv.2212.13460,Matos-Shabani}, 
there is a continued theoretical debate of its underlying theoretical origin. A common scenario invokes simultaneous breaking of space-inversion, accompanied by the spin-orbit coupling (SOC), and time-reversal 
symmetries~\cite{Edelstein1989,Edelstein1996,Dimitrova2007,PhysRevLett.102.227005,Mineev2012,
Daido2021,Yuan2021,Smith2021,He2022,Scammell2022,
Ilic2022,Davydova2022,arxiv.2302.04277,Baumgartner2022,BaumgartnerSI2022,Fuchs2022,arxiv.2212.13460,Matos-Shabani}.  
However, there are  
interesting alternatives suggesting that the nonreciprocal supercurrent originates from inhomogeneous edge transport and stray fields, or from purely orbital effects involving diamagnetic currents,  
or limitations from asymmetric vortex-protrusions, all without a need for SOC and Zeeman-related phenomena~\cite{arxiv.2205.09276,arxiv.2301.01881,arxiv.2207.03633,Suri2022}. 

In this work we present a unified origin of the SDE in different platforms by generalizing phenomenological Ginzburg-Landau (GL) 
theory~\cite{Edelstein1989,Mineev1994,Edelstein1995,Edelstein1996,Samokhin2004,MineevPRB2008,Mineev2012,Agterberg2012,Fuchs2022}. 
With the self-consistent solution of the noncentrosymmetric quasi-two-dimensional (2D) superconductors and superconductor/normal region/ superconductor (S/N/S) JJs, we obtain an excellent agreement with the observed SDE and current-phase relation 
(CPR)~\cite{Ando2020,Baumgartner2022,BaumgartnerSI2022,Jeon2022,Turini2022,arxiv.2212.13460}. 
We further predict overlooked trends for the SDE magnitude with the sample geometry.


\begin{figure*}
 \includegraphics[width=.84 \linewidth]{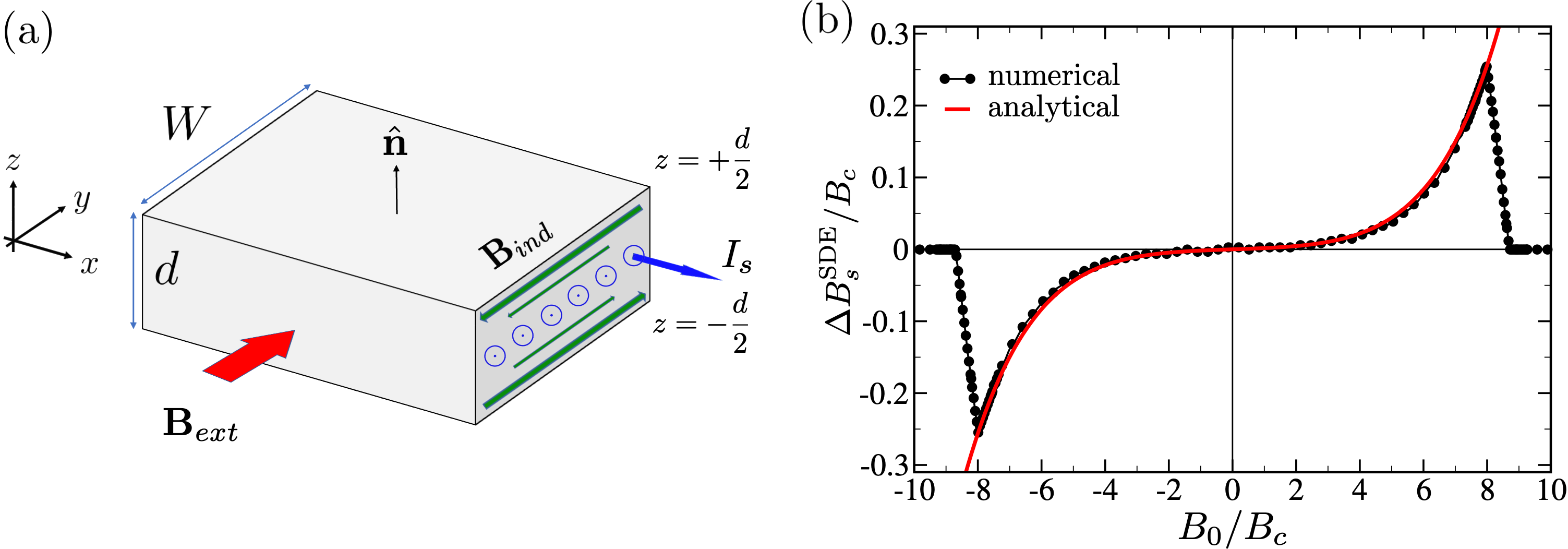}
 \caption{(a) Schematic of a noncentrosymmetric superconducting film of thickness $d$, in a constant external in-plane magnetic field, $\mathbf{B}_{ext}=B_0\,\hat{y}$, carrying a steady DC supercurrent, $I_s$, which generates an induced magnetic field, $\mathbf{B}_{ind}(z)$, inside the slab (green arrows).
 (b) SDE versus  
 in-plane field, $B_0$, (in units of the thermodynamic field $B_c$), for a quasi-2D film. 
 Black symbols: the full numerical solution of 
 Eqs.~(\ref{Eq:GL2-aver})--(\ref{Eq:GL1-aver}). 
 Red curve: the analytical result given by
 Eq.~(\ref{Eq:SDE-current}) valid in the limit $(d/\lambda)^2\ll 1$, $(d/\ell)^2\ll 1$, and $B_I/B_{c}\ll 1$, with $\ell$  
 from Eq.~(\ref{eq:LE}). 
 The SDE supercurrent density 
 $\Delta j_s^{\text{SDE}}$ is converted to the corresponding SDE  
 field difference, $\Delta B_s^{\text{SDE}}=\mu_0\Delta j_s^{\text{SDE}}d/2$, normalized to $B_{c}$. 
 For calculations we used: $\xi/\lambda=1.3$, $\ell/\lambda=3$, and $d/\lambda=0.5$.}
 \label{Fig1}
\end{figure*}

Starting from the microscopic Bogoliubov--de Gennes equations in the presence of (i) isotropic Rashba SOC
of strength $\alpha_\mathrm{R}$, with the SOC-defined unit vector normal to the film, $\hat{\mathbf{n}}$,
and (ii) 
the magnetic field 
$\mathbf{B}$, which yields
an orbital coupling and Zeeman interaction, defined by g-factor, $g$,  
one systematically 
derives the generalized GL free-energy density for the condensate wave function 
$\psi$~\cite{Edelstein1989,Mineev1994,Edelstein1995,Edelstein1996,Samokhin2004,MineevPRB2008,Mineev2012} 
\begin{equation}\label{Eq:GL-functional-plus-Lifshitz}
    F=a|\psi|^2+\frac{b}{2}|\psi|^4+\frac{|\mathbf{D}\psi|^2}{4m}+\frac{\mathbf{B}^2}{2\mu_0}
    -\frac{\mathcal{K}}{2} (\hat{\mathbf{n}}\times\mathbf{B})\cdot \mathbf{Y}_\psi.
\end{equation}
The first four standard terms are parameterized  
by  
coefficients 
$a$, $b$, effective 
mass $m$, and the covariant derivative $\mathbf{D}=-i\hbar\mathbf{\nabla}-2e\mathbf{A}$, with the electron charge $e<0$, and vector potential $\mathbf{A}$.
The last term, $\propto 
\mathbf{Y}_\psi=(\psi)^*\mathbf{D}\psi+\psi(\mathbf{D}\psi)^*$, is 
the (isotropic) 
\textit{Lifshitz invariant}~(LI)~\cite{Lifshitz1-JETP1941,Lifshitz2-JETP1941},  
a figure of merit of the noncentrosymmetry. 
LI is responsible for the spatial modulation of the order parameter
$\psi=e^{i\mathbf{k}_\mathcal{K}\cdot\mathbf{r}}|\psi|$, the helical phase, where the wave vector, 
$\mathbf{k}_\mathcal{K}\propto 2m\mathcal{K}(\hat{\mathbf{n}}\times\mathbf{B})/\hbar$, with   
the coupling constant  
$\mathcal{K}\simeq 3g \mu_\mathrm{B} \alpha_\mathrm{R}/(\hbar E_\mathrm{F})$~\cite{Edelstein1996}\footnote{The coupling constant $\mathcal{K}$ is also temperature, $T$, dependent, the full expression is  
$\mathcal{K}=3\frac{\alpha_R}{\hbar}\frac{g\,\mu_{B}}{v_F\,p_F}\,f_3\left(\frac{\alpha_R\,p_F}{\hbar\,\pi\, k_{\mathrm{B}} T}\right)$, where
$f_3(x)\simeq 0.475\int\limits_{0}^{\pi} \mathrm{d}t \sum\limits_{n=0}^{\infty}\frac{\sin{t}\,(x\sin{t})^2}{(2n+1)^3[(2n+1)^2+(x\sin{t})^2]}$. Approximation in the main text holds for low $T$.},
of the dimension $\text{m}\cdot\text{C/kg}$,
where $E_\mathrm{F}$ is  
the Fermi energy.  
We use common coherence and penetration lengths, $\xi$, $\lambda$, and the thermodynamical critical field, $B_{c}=\Phi_0/(2\sqrt{2}\pi\lambda\xi)$, where $\Phi_0$ is the magnetic flux quantum. 
The presence of LI yields an additional scale, the  
\textit{Lifshitz-Edelstein length}~\cite{Fuchs2022} 
\begin{equation}
\ell = b/({2\mathcal{K}\mu_0 |e| |a|}). 
\label{eq:LE}
\end{equation} 
We will use $\xi$, $\lambda$, $\ell$, and $B_c$ to express our final results.

After introducing 
the \textit{shifted momentum operator} 
$\mathbf{D}_{\mathcal{K}}=\mathbf{D}-\hbar\mathbf{k}_\mathcal{K}$,
and the \textit{shifted $a$-coefficient}, 
$a_\mathcal{K}=a-(\hbar\mathbf{k}_\mathcal{K})^2/(4m)$, 
by subtracting, correspondingly, the center-of-mass momentum and energy of the helical 
Cooper pairs, we obtain the  
GL~equations
\footnote{At the interface to vacuum, from the variation of $F$, one obtains the boundary condition
$0=(\hat{\boldsymbol{\nu}}_{\text{out}}\cdot\mathbf{D}_\mathcal{K})\psi\bigl|_{\text{interface}}$, 
where $\hat{\boldsymbol{\nu}}_{\text{out}}$ is the unit normal vector to the interface pointing into vacuum. In this work, all condensate wave-functions satisfy the above boundary conditions.}
\begin{align}\label{Eq:GL-Eqs}
    0 
    =
    \frac{\mathbf{D}_\mathcal{K}^2}{4m}\psi
    +a_\mathcal{K}\psi + b|\psi|^2\psi,
    \ \  
    \ \
    \mathbf{j}_\mathrm{s}
    =
    \boldsymbol{\nabla} \times \,\mathbf{H},
\end{align}
where, $\mathbf{H} = \mathbf{B}/\mu_0 + \tfrac{1}{2}\mathcal{K}\, (\hat{\mathbf{n}}\times\mathbf{Y}_\psi)$, and the supercurrent density  
is  
\begin{align}
     \mathbf{j}_\mathrm{s} 
     &= 
     \frac{e}{2m}\mathbf{Y}_\psi-2e\mathcal{K}|\psi|^2 (\hat{\mathbf{n}}\times\mathbf{B})\label{Eq:supercurrent}.
\end{align}

\paragraph*{\textbf{SDE in thin noncentrosymmetric films:}} We first 
consider a quasi-2D film, long and wide ($W$) 
compared to $\xi$, with 
 $\hat{\mathbf{n}}=\hat{z}$, as shown in Fig.~\ref{Fig1}(a). We assume the film thickness 
 $d \lesssim \xi$,  
implying that $\psi$ is approximately $z$-independent. 
The quasi-2D film  
is in the  
constant external (laboratory) in-plane  
field, $\mathbf{B}_{ext}=B_0\,\hat{y}$, and is driven by a steady supercurrent, $I_s$, with the corresponding density, $I_s/(W\,d)$.
As a consequence of the supercurrent drive,  
inside the film,  
$z\in[-d/2, +d/2]$, there is an induced  
field, $\mathbf{B}_{ind}(z)$. Therefore, the
total resulting 
field inside 
the slab, $\mathbf{B}=\mathbf{B}_{ext}+\mathbf{B}_{ind}(z)$, is spatially inhomogeneous, and the LI cannot be removed by a gauge transformation from the 
quasi-2D free-energy $F$.  This is in contrast to a spatially-homogeneous field treated 
in Refs.~\cite{Agterberg2012,Smidman2017}. 
Such a subtle point is sometimes overlooked and  
the removal of the LI by a gauge transformation is erroneously used beyond  
its applicability.

Given the above assumptions, we look for  
$\psi$ and  
$\mathbf{B}_{ind}$, inside the film,  
in the forms: 
$\psi(x)=e^{ikx}\,f\sqrt{|a|/b}$ and $\mathbf{B}_{ind}(z)=(-2 B_I z/d)\,\hat{y}$,
where the unknown real factors, the helical wave vector, $k$, the dimensionless wave function, $f>0$, and the induced magnetic field, $B_I$, will be determined later. 
Within the Coulomb gauge, the corresponding 
$\mathbf{A}=A(z)\,\hat{x}=\left(B_0 z-B_I z^2/{d}\right)\,\hat{x}$,  
satisfies $\boldsymbol{\nabla} \times\,\mathbf{A}=\mathbf{B}_{ext}+\mathbf{B}_{ind}(z)$. By substituting  
the above  
$\psi(x)$ and $\mathbf{B}_{ind}(z)$
into the corresponding GL  
Eq.~(\ref{Eq:GL-Eqs}), and  
$\mathbf{j}_s=j_s(z)\,\hat{x}$, Eq.~(\ref{Eq:supercurrent}), one obtains a nonlinear system of differential equations (not explicitly shown here for brevity), which couples $k$, $f$, and $B_I$ with $A(z)$, $A^2(z)$ (diamagnetic terms), and their various $z$-derivatives.

In superconducting thin films~\cite{schmidt1997physics}, 
the homogeneity of $\psi$ along the $z$-direction allows one to average the underlying system of differential equations over the thickness.
We replace $j_s(z)$, and $A(z)$, and its various powers and $z$-derivatives, by their averaged values $\langle j_s\rangle=I_s/(W\, d)$, $\langle A\rangle=-B_{I}{d}/{12}$, etc., 
where  
$\langle O\rangle \equiv \int_{-d/2}^{+d/2} \mathrm{d}z\,O(z)/d$.
This averaging removes $\mathbf{A}$-dependent terms and converts the system of differential equations into an algebraic one
\begin{align}
    k
    &=
   -\frac{\sqrt{2}}{\xi}\frac{B_{I}}{B_{c}}\frac{\lambda}{d}\left[\frac{1}{f^{2}}-\frac{1}{24}\frac{d^2}{\lambda^2}\right],
    \label{Eq:GL2-aver}
    \\[3pt]
    B_I
    &=
    \mu_0 \langle j_s\rangle\frac{d}{2}+f^2\frac{d}{2\ell} B_0
    =
    \mu_0 \frac{I_s}{2W}+f^2\frac{d}{2\ell} B_0,\label{Eq:GL3-aver}
    \\[3pt]
    0
    &= 
    \xi^2 k^2 - \frac{\xi k }{6\sqrt{2}}\frac{B_{I}}{B_{c}}\frac{d}{\lambda} + 
    \left(\frac{1}{24}\frac{B_0^2}{B_{c}^2}+\frac{1}{160}\frac{B_I^2}{B_{c}^2}\right)\frac{d^2}{\lambda^2}\nonumber\\ 
    &-1 + f^2
    +\xi k\,\frac{\sqrt{2}\lambda}{\ell}\,\frac{B_0}{B_{c}} -\frac{1}{4} \frac{B_0}{B_{c}}\frac{B_I}{B_{c}}\frac{d}{\ell}\label{Eq:GL1-aver}. 
\end{align}
Apart from the averaging, the above system of equations for $k$, $f$, and $B_I$, 
in terms of experimentally-defined 
$B_0$ and $I_s$, becomes exact.
\begin{figure*}
 \includegraphics[width=.84 \linewidth]{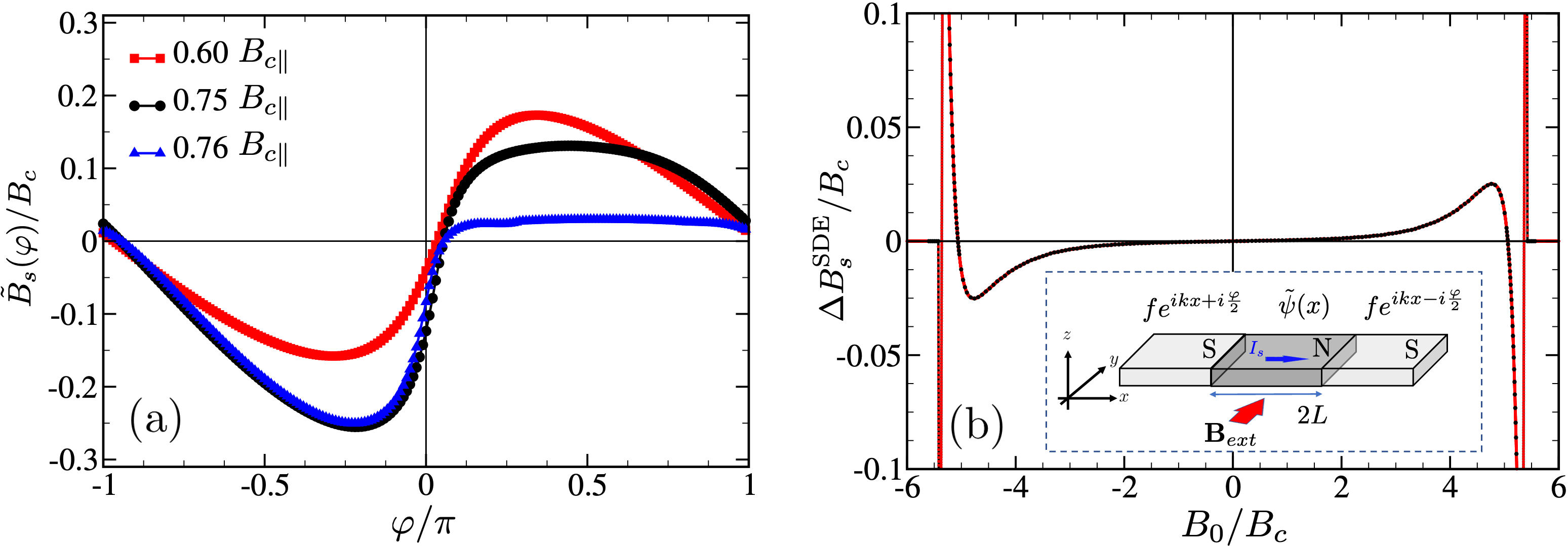}
 \caption{
 CPR and SDE for a long noncentrosymmetric SNS JJ with the same parameters in the N and S regions, i.e.,~$\xi=\tilde{\xi}$, $\lambda=\tilde{\lambda}$, $\ell=\tilde{\ell}$, $d=\tilde{d}$, and $m=\tilde{m}$:
 (a) Distorted CPRs,  
 for different external  
 fields: $0.6$ (red), $0.75$ (black), and $0.76$ (blue) multiples of parallel-critical-field,
 $B_{c\parallel}$, shown  
 $\tilde{B}_{s}(\varphi)=\mu_0 \tilde{j}_{s}(\varphi) \tilde{d}/2$, normalized to $B_{c}=\tilde{B}_{c}$,  
 as a function of $\varphi/\pi$.  
 (b) SDE versus the in-plane field $B_0$, normalized to $B_{c}$. The supercurrent difference $\Delta\tilde{j}_s^{\text{SDE}}$ is converted to the corresponding field difference $\Delta\tilde{B}_s^{\text{SDE}}=\mu_0 \Delta\tilde{j}_s^{\text{SDE}} \tilde{d}/2$ and is plotted in units of $B_{c}$. 
 The black symbols: computed data points, red curve: an interpolation line.
 Inset: JJ geometry  
 with the phase difference $-\varphi+2kL$. 
 Here, $\xi/\lambda=1.3$, $\ell/\lambda=3$, $d/\lambda=0.7$, and $L/\lambda=1.95$.}
 \label{Fig2}
\end{figure*}
The nonlinearity and algebraic complexity preclude writing the closed-form analytical solutions of 
Eqs.~(\ref{Eq:GL1-aver})-(\ref{Eq:GL2-aver}).
However, from Eq.~(\ref{Eq:GL1-aver}) it is clear 
that the \textit{magnetochiral term} $\propto 
B_0 B_I d/\ell \sim B_0 \langle j_s \rangle d/\ell$ gives the 
condensate wave function $f$, and hence the maximal supercurrent $j_s^{max}$ the condensate can carry, 
a nontrivial dependence on the mutual orientations of $\mathbf{B}_{ext}$ and $\mathbf{j}_s$. Furthermore, 
the strength of the magnetochiral effect  
is also driven by the thickness $d$ and the product of the Rashba SOC strength 
and g-factor, $1/\ell\simeq g\alpha_\mathrm{R}$, recall Eq.~(\ref{eq:LE}). 

To derive some analytical results, we  
systematically  
keep  
only linear and quadratic terms in the small parameters $d/\lambda$ and $d/\ell$, as well as linear terms  
in $B_I/B_{c}$. For thin films, $B_0$ can be 
up to parallel-critical-field $B_{c\parallel}=2\sqrt{6} B_{c}\lambda/d\gg B_{c}$, and therefore, we do not employ any restriction on~$B_0/B_{c}$. 
While these approximations simplify the system of algebraic equations,  
their closed solutions are still too cumbersome to be written explicitly. 
One can show that by 
keeping $B_0$ fixed and varying $\langle j_s\rangle$ compared to 0. 
The physical solutions possess real-valued $k$ and positive-valued $f^2$. However, there exists positive and negative maximal supercurrent densities $j_{s>}^\mathrm{max}>0$ and $j_{s<}^\mathrm{max}<0$ that when overcome by $\langle j_s\rangle$, the physical solutions no longer exist. At these points the system undergoes the first-order phase transition and jumps from the superconducting into the normal phase. While the expressions for $j_{s>}^\mathrm{max}$ and $j_{s<}^\mathrm{max}$ remain 
complex,
the SDE difference $\Delta j_s^{\text{SDE}}=j_{s>}^\mathrm{max}-|j_{s<}^\mathrm{max}|$ , within the considered  
approximations, has a compact form~\footnote{The corresponding SDE efficiency 
$\eta_s^{\text{SDE}}=(j_{s>}^{max}-|j_{s<}^{max}|)/(j_{s>}^{max}+|j_{s<}^{max}|)$ within the
same level of approximation equals $\eta_s^{\text{SDE}}
    \simeq
    \frac{\sqrt{3}}{36}\frac{d^2}{\lambda^2}\left(\frac{B_0}{B_{cm}}\frac{\lambda}{\ell}\right)\left[2+\left(\frac{B_0}{B_{cm}}\,\frac{\lambda}{\ell}\right)^2\right]^{\frac{1}{2}}$.}
\begin{align}\label{Eq:SDE-current}
\Delta j_s^{\text{SDE}}
&=
\frac{B_{c}\,d^2}{54\mu_0\lambda}\left(\frac{B_0}{B_{c}}\frac{\lambda}{\ell}\right)\left[2+\left(\frac{B_0}{B_{c}}\frac{\lambda}{\ell}\right)^2\right]^2.
\end{align}
We compare the accuracy of this approximate SDE result in Eq.~(\ref{Eq:SDE-current}) with the full numerical solution of Eqs.~(\ref{Eq:GL2-aver})--(\ref{Eq:GL1-aver}), without any approximations, in Fig.~\ref{Fig1}(b). The deviations at higher fields stem from the employed approximation which neglects some higher-order terms.

We observe that in quasi-2D films $\Delta j_s^{\text{SDE}} \propto d^2$, i.e.,~the SDE effect based on the lowest-order LI vanishes as $d\rightarrow 0$,  
as expected from the assumption that LI could be removed by a gauge transformation.  
An analog of the higher-order LI that works for a pure 2D system was proposed  
in Ref.~\cite{He2022}. 
Furthermore,  
the derivative of $\Delta j_s^{\text{SDE}}$ with respect to the external field  
is 
\begin{equation}\label{Eq:SDE-current-2}
        \left.\frac{\mathrm{d}}{\mathrm{d} B_0}\Delta j_s^{\text{SDE}}\right|_{B_0=0} 
        = \frac{2}{27 \mu_0}\frac{1}{\ell}\frac{d^2}{\lambda^2}
        = \frac{1}{18 \mu_0}\frac{g\alpha_\mathrm{R}}{E_\mathrm{F}}\frac{d^2}{\lambda^4} ,
\end{equation}
which allows one to extract $1/\ell = (3/4)g\alpha_\mathrm{R}/(E_\mathrm{F}\lambda^2)$, and hence the ratio of the Rashba SOC and Fermi energy.
Thus the SOC characteristics of quasi-2D films can be  
probed by the SDE  
in  
the superconducting phase.


\paragraph*{\textbf{SDE and CPR in SNS Josephson junctions:}} As the second example, we consider a long ballistic SNS JJ consisting of noncentrosymmetric S and N elements 
and look for a solution $\tilde{\psi}$ of Eqs.~(\ref{Eq:GL-Eqs}) in the $2L$-long N-region; see the inset in Fig.~\ref{Fig2}. The junction is placed in an in-plane field $\mathbf{B}_{ext}=B_0\,\hat{y}$ and carries a steady supercurrent $I_s$; for the induced field $\mathbf{B}_{ind}$ and the vector potential $\mathbf{A}$ we employ the same ansatz and Coulomb
gauge 
as before. 
Requiring the continuity of the condensate wave functions, we match $\tilde{\psi}(x)$ at $x=\pm L$ with  
$\psi(x)=f e^{ik x\pm i\varphi/2}$, in the sufficiently long and wide identical left and right superconductors, S.
The total phase difference between their ends is $-\varphi+2kL$, see the 
inset in Fig.~\ref{Fig2}. The external phase $\varphi$ can be controlled, e.g.,~by a 
flux loop. 
To distinguish the parameters and quantities in the S and N regions, we use the tilde symbols for those 
associated with N, for example, $\tilde{a}>0$, while $a<0$.

Following a similar procedure for a thin film, we transversely average GL Eqs.~(\ref{Eq:GL-Eqs})~and~(\ref{Eq:supercurrent})
as well as linearize them (ignoring the cubic term in $\tilde{\psi}$), since the JJ is long and the proximity-induced 
$\tilde{\psi}$ is suppressed.
The resulting equation for $\tilde{\psi}$, including the diamagnetic terms, is 
\begin{align}\label{Eq:GL1-NormalState}
    0 
    &=
    \frac{\mathrm{d}^2\tilde{\psi}}{\mathrm{d}x^2}
    -2i\underbrace{
    \frac{1}{\sqrt{2}\tilde{\xi}}\left[
    \frac{1}{12}\frac{\tilde{d}}{\tilde{\lambda}}\frac{\tilde{B}_{I}}{\tilde{B}_{c}}-\frac{\tilde{\lambda}}{\tilde{\ell}}\frac{B_0}{\tilde{B}_{c}}
    \right]}_{\tilde{\varepsilon}}\frac{\mathrm{d}\tilde{\psi}}{\mathrm{d}x}\\
    &-
    \underbrace{\frac{1}{\tilde{\xi}^2}\left[
    1-\frac{\tilde{d}}{4\tilde{\ell}}\frac{B_0 \tilde{B}_{I}}{\tilde{B}_{c}^2}+
    \frac{\tilde{d}^2}{\tilde{\lambda}^2}\left(\frac{B_0^2}{24 \tilde{B}_{c}^2}+\frac{\tilde{B}_{I}^2}{160 \tilde{B}_{c}^2}\right)
    \right]}_{\tilde{\alpha}>0}\tilde{\psi},  \nonumber
\end{align}
where due to the linearization, $\tilde{B}_{I}=\mu_0 \langle \tilde{j}_{s}\rangle \tilde{d}/2$.
The solution of the above equation is a superposition of $e^{i\tilde{\varepsilon}x+\tilde{q}x}$ and $e^{i\tilde{\varepsilon}x-\tilde{q}x}$, where, 
$\tilde{q}=\sqrt{\tilde{\alpha}-\tilde{\varepsilon}^2}$. Matching it with the wave functions 
in left and right S, one gets $\tilde{\psi}(x)$ and, according to Eq.~(\ref{Eq:supercurrent}), the supercurrent density
\begin{align}
  \tilde{j}_{s}(\varphi)
  &=
  \sqrt{8}\tilde{q}f^2\frac{m}{\tilde{m}}\frac{B_{c}}{\mu_0}
  \frac{\xi}{\lambda}\,\sin{[(\varphi-2kL)+2\tilde{\varepsilon}L]}
  e^{-2\tilde{q} L},\label{Eq:SNS-CPR}
\end{align}
where $f$, $k$, $\xi$, $\lambda$, and $B_{c}$ correspond to the two S regions, while the ratio $m/\tilde{m}$ accounts for different effective masses (density of states) in the S and N.

Our expression Eq.~(\ref{Eq:SNS-CPR}) is a generalization of the result for anomalous phase shift 
$\phi_0=-4L\tilde{\mathcal{K}}\tilde{m}B_0/\hbar$  
from Ref.~\cite{Buzdin2008} for the noncentrosymmetric SNS junction, but with (i) considering the LI only inside the N region and (ii) a calculation limited by including only $B_0/\tilde{B}_c$ term, ignoring its higher powers and also $\tilde{B}_{I}/\tilde{B}_c$ terms.
Unlike this result, which yields a simple $\phi_0$-shift in the harmonic CPR and no SDE~\cite{amundsen2022colloquium},  
our  
Eq.~(\ref{Eq:SNS-CPR}), as we discussed below, reveals several key differences in $\tilde{j}_s(\varphi)$, including 
robust SDE and strong anharmonicity, characteristic also for JJs revealing topological superconductivity~\cite{PhysRevLett.126.036802}.  
All magnetochiral terms $\propto 1/\tilde{\ell}$ and those induced by $\tilde{j}_s(\varphi)\sim\tilde{B}_{I}$ itself, give rise to the dynamically-generated CPR stemming from the self-consistent problem, $\tilde{j}_s(\varphi)=\mathbb{F}[\tilde{j}_s(\varphi)]$:
\textbf{(1)} Having a supercurrent $I_s$ passing through the SNS~junction, we calculate the supercurrent densities $\langle j_s\rangle =I_s/(W d)$ and $\langle \tilde{j}_{s}\rangle =I_s/(\tilde{W} \tilde{d})$ in the S and N regions;
\textbf{(2)} Obtained 
$\langle j_s\rangle $ in the S~region defines, for a given external field $B_0$, the dimensionless wave functions, $f$, wave vector, $k$, and
induced field, $B_I$, see Eqs.(\ref{Eq:GL2-aver})--(\ref{Eq:GL1-aver}); 
\textbf{(3)} Knowing $\langle \tilde{j}_{s}\rangle $ specifies $\tilde{B}_{I}$, $\tilde{\varepsilon}$, and $\tilde{q}$, which, along with the values of $f$ and $k$, specify the CPR $\tilde{j}_{s}(\varphi)$ according to 
Eq.~(\ref{Eq:SNS-CPR});
\textbf{(4)} The phase difference $\varphi_{I_s}$ corresponding to the given supercurrent 
$I_s$ should be chosen such that 
$\tilde{j}_{s}(\varphi_{I_s})=I_s/(\tilde{W} \tilde{d})=\langle \tilde{j}_{s}\rangle $.

The results of the above self-consistent procedure are displayed in Fig.~\ref{Fig2}(a), which shows the anharmonic CPR for the fully symmetric SNS junction with 
$L/\lambda=1.95$, $d/\lambda=0.7$, and several in-plane fields. The pronounced harmonic 
distortion of the CPR when evolving the phase $\varphi$ gives, for different external fields, different magnitudes of positive, $\tilde{j}_{s>}^\mathrm{max}$, and negative, $\tilde{j}_{s<}^\mathrm{max}$, critical currents, i.e., the JJ SDE. 
We plot the corresponding SDE difference $\Delta\tilde{j}_s^{\text{SDE}}=\tilde{j}_{s>}^\mathrm{max}-|\tilde{j}_{s<}^\mathrm{max}|$ with the in-plane 
$B_0$ in Fig.~\ref{Fig2}(b). For positive $B_0$,  with its increase,  
$\Delta\tilde{j}_s^{\text{SDE}}$ gradually grows, first linearly, and then with admixed cubic and quintic dependence on $B_0$. 
However, at a certain field range ($B_0/B_{c}\sim 5$), the 
SDE starts to saturate, then decreasing, crossing to negative values and again upturning and returning to zero. Numerical data in the sign-changed transition region are showing a certain numerical instability, e.g.~CPR does not exist for all phases or becomes multi-valued, 
therefore we plot the reliable data 
points by black symbols, and their interpolation, 
including the unstable region, with red line.


\paragraph*{\textbf{Conclusions:}}

Our theoretical framework offers a transparent and unified approach to analyze the SDE in different structures and use it to probe the SOC in quasi-2D systems. Surprisingly, a self-consistent solution to simple algebraic equations already provides an important tool to examine the magnetochiral properties of JJs, including the anomalous phase shift, anharmonic CPR, and the sign reversal of the SDE at high magnetic fields.
These signatures are also important in the studies of topological superconductivity~\cite{PhysRevLett.126.036802,Matos-Shabani}, while the spin-triplet proximity-induced
superconductivity accompanying the SDE in noncentrosymmetric systems is directly relevant for superconducting spintronics~\cite{amundsen2022colloquium}.

While we have focused on a commonly assumed SOC linear in the wave vector, our approach could be also generalized to consider an anisotropic LI~\cite{Samokhin2004} inherent to JJs with cubic 
SOC~\cite{Alidoust:PhysRevB2021,PhysRevB.107.035435} 
and look also for its a non-linear and even multi-component order-parameter 
generalizations~\cite{Mineev1994,Samokhin2014}. 
The resulting proximity-induced $f$-wave 
superconductivity~\cite{Alidoust:PhysRevB2021}, 
along with including the nonlinear Meissner 
effect~\cite{PhysRevB.51.16233,PhysRevLett.82.3132,PhysRevLett.110.087002,PhysRevB.56.11279,PhysRevB.58.8738} 
provides unexplored directions for SDE studies and their implications for unconventional superconductivity. 
With the AC applied magnetic field, the generation of higher harmonics~\cite{Alidoust:PhysRevB2021,fukaya2022anomalous} and their anisotropy could be used to measure the resulting nonlinear~\cite{PhysRevLett.82.3132,PhysRevB.58.8738} and magnetochiral contributions. 
           


\paragraph*{\textbf{Acknowledgements:}}
We thank C.~Strunk, N.~Paradiso, J.~Fabian, D.~Agterberg, L.~Rokhinson, A.~Buzdin, M.~Milo\v{s}evi\'{c}, and A.~Vagov for useful discussions. 
D.K.~acknowledges a partial support from the IMPULZ project IM-2021-26---SUPERSPIN funded by the Slovak Academy of Sciences, VEGA Grant No.~2/0156/22---QuaSiModo, and from the COST Action CA21144---SUPERQUMAP.
Work in Regensburg was partly supported by Deutsche Forschungsge\-meinschaft (DFG, German Research Foundation) within Project-ID 314695032---SFB 1277~(projects A07 and B07) and Project-ID 454646522---``Spin and magnetic properties of superconducting tunnel junctions''. 
I.\v{Z}. was supported by
the U.S.~ONR through Grants No.~N000141712793 and MURI No.~N000142212764, and
NSF~Grant~No.~ECCS-2130845.

 
\bibliography{library}

\end{document}